\documentstyle[sprocl]{article}

\def\Journal#1#2#3#4{{#1} {\bf #2}, #3 (#4)}

\def\NPB{{\em Nucl. Phys.} B}

\def\PRL{\em Phys. Rev. Lett.}

\def\be{\begin{equation}}
\def\ee{\end{equation}}
\def\bea{\begin{eqnarray}}
\def\eea{\end{eqnarray}}

\begin{document}
\title{ 
A SYSTEMATIC APPROACH TO ANOMALOUS 
PHENOMENA AT HIGH ENERGIES}
\author{MIKUL\' A\v S BLA\v ZEK}
\address{Institute of Physics, Slovak Academy of Sciences,
84228 Bratislava,\\ Slovak Republic\\ E-mail: blazek@savba.sk}

\maketitle\abstracts{ 
In one-dimensional case the search for  presence of the anomalous
phenomena in multiplicity distributions is usually performed  in frame 
of the  horizontal, vertical and mixed types of the analysis.
  We show that if the data involve a d-dimensional phase space, there 
exists a convenient procedure, generalizing the one-dimensional approach,
which allows to introduce more non-trivial types of the
analysis; we formulate them in some detail in one-, two-,  and partially 
also in three-dimensions.}

\section{Introduction}
  Search for anomalous phenomena at high energies is usually performed
in terms of a convenient kind of  statistical moments, the information 
about the presence of those phenomena being encoded in appropriate sort  
of the scaling indices.
 Quite often, the one-dimensional cases are considered${^1}$ and usually 
three types of the analysis are applied there${^2}$, namely, 
the {\em horizontal} $(H)$, {\em vertical} $(V)$, and the {\em
mixed} $(HV)$ ones.
  However,
there is still missed  a systematic approach  to higher dimensions
as well as a deeper  rooted physical theory or at least a more or
less reliable  dynamical model for the scaling indices
characterizing those phenomena.

    Let us add that in the cases  considered in the present contribution,
the appearance of the  non-statistical, anomalous fluctuations, induces
the need to take into account the
non-continuous statistical distributions.

      In principle, the notion ``anomalous'' means (special, see below) 
``non-statistical'' while ``statistical'' denotes ``pure Poissonian''.

When investigating the presence and properties of the anomalous phenomena
in high energy physics, the factorial moments of rank $q$ are usually applied,
their normalization being such that for the pure Poissonian distribution
they are equal to unity (for $q$= 1, 2, 3, ...).
This means that their deviation of unity (at least for some values of the 
rank $q$) signifies the presence of the non-statistical behavior in the
corresponding data.

 In the present contribution a sample of $E$ events is
investigated, the counting index of the events being denoted by $e$,
i.e. $e$=1, 2, 3, ..., $E$. 
 The number of (charged) particles observed in a given region of the
phase-space under consideration is denoted by $n$. In what follows
also the factorial multinomial, $\cal F$, introduced by
\be
 {\cal F} ~=~ {\cal F}(n;q) \equiv~ n(n-1) \ldots (n-q+1)~,
\ee 
is applied.
                      
\section{One-dimensional case}
We assume that the dependence of the multiplicity distribution of
(charged, produced) secondaries on one variable (say, 
the pseudorapidity ${\eta}$ of
every particle under consideration) is known from the  observed
collisions.

    Let the multiplicity data be known in a pseudorapidity 
window $\Delta \eta$, and, eventually, this window
be partitioned into $M$ equal size bins, $\delta \eta, 
\delta \eta = (\Delta \eta)/M$; the index counting the bins is denoted
by $m$, i.e. $m = 1,2,3, \ldots, M.$ It is understood that the
number of particles  in every elementary cell specified by two numbers,
namely, $(me)$, i.e. $n_{me}$, is deduced from the experimental
output.

   Usually it is said that the anomalous phenomenon (intermittency)
is present if the experimental data exhibit (with increasing number of
bins, $M$) a linear dependence between the logarithm of the factorial
moments and logarithm of the number of bins, $M$, i.e., if the factorial
moments reveal a singularity of the form $M^{a_q}$, $a_q$ being 
non-vanishing quantities; they are called slopes or
 scaling indices${^3}$. It is worthwile to introduce also 
the following notations: to emphasize the {\bf I}ndividual, 
{\bf I}ndependent averaging of the expression entering the numerator
{\em or} denominator of the corresponding moment,
$$  {\bf I}_m \equiv \frac{1}{M} \sum_{m=1}^M~,~~~ {\bf I}_e 
\equiv \frac{1}{E} \sum_{e=1}^E~, $$
and, to emphasize the averaging over {\bf B}oth, the numerator {\em as
well as} the denominator,
$$ {\bf  B}_m ~\equiv~ \frac{1}{M} \sum_{m=1}^M~, ~~~ {\bf B}_e 
\equiv~ \frac{1}{E} \sum_{e=1}^E~. $$

Now, application of the  {\em horizontal} type of the analysis means that
the factorial moments in the $(me)-$th cell  
  $[F(q;M)]_{me}^{(H)}$ ,
   in the $e-$th event  
$[F(q;M)]_e^{(H)}$, 
    and in the whole sample
$[F(q;M)]^{(H)}$, 
    are introduced, respectively, by
\be
 [F(q;M)]_{me}^{(H)}~=~ \frac{{\cal F}(n=n_{me};q)}{[{\bf I}_{m'}
n_{m' e}]^q}~       
\ee
and
\bea  
[F(q;M)]_e^{(H)} ~&=&~ {\bf I}_m  [F(q;M)]_{me}^{(H)}  \nonumber \\
       &=& \frac{{\bf I}_m {\cal F}(n=n_{me};q)}{[{\bf I}_{m'}
n_{m' e}]^q}~,
\eea
\be
[F(q;M)]^{(H)} ~=~ {\bf B}_e  [F(q;M)]_e^{(H)}
\ee
where ${\cal F}$ is given by Eq. 
(1). In this case the scaling properties
are investigated by means of the relation,
\be
[F(q;M)]^{(H)} ~ \propto~  f_q^{(H)} M^{a_q^{(H)}}~.
\ee

 The {\em vertical} type of the analysis  is specified by the following
form of the factorial moments, 
\be
 [F(q;M)]_{me}^{(V)}~=~ \frac{{\cal F}(n=n_{me};q)}{[{\bf I}_{e'}
n_{m e'}]^q}~       
\ee
and
\be  
[F(q;M)]_m^{(V)} ~=~ {\bf I}_e  [F(q;M)]_{me}^{(V)} 
\ee
\bea
[F(q;M)]^{(V)} ~&=&~ {\bf B}_m  [F(q;M)]_m^{(V)}~  \nonumber    \\
                &=& ~ {\bf B}_m {\bf I}_e [F(q;M)]_{me}^{(V)} ~
\eea
where ${\cal F}$ is still given by (1). Now, the scaling property is
formulated in the following way,
\be
[F(q;M)]^{(V)} ~ \propto~  f_q^{(V)} M^{a_q^{(V)}}~.
\ee

The {\em mixed} type of the analysis is characterized by the following
expressions,
\be
[F(q;M)]_{me}^{(HV)} ~=~  \frac{{\cal F}(n=n_{me};q)}{[{\bf I}_{m'}
{\bf I}_{e'} n_{m' e'}]^q}
\ee
and
\be
[F(q;M)]^{(HV)}~ =~  {\bf I}_m {\bf I}_e [F(q;M)]_{me}^{(HV)} 
   \propto f_q^{(HV)} M^{a_q^{(HV)}}~.
\ee

 We conclude this section by the observation that the expressions
$$ {\bf B}_e{\bf I}_m~,~~~{\bf B}_m{\bf I}_e~ , ~~~ {\bf I}_m {\bf I}_e
 $$
represent, respectively, a shorthand notation of the three types of the
analysis mentioned above. 
It is perhaps worth to mention that in case of the silicon 
(at 14.6 A GeV/c)
colliding with the emulsion nuclei (the BNL data),
those three types of the analysis lead${^4}$ to similar values of the 
slopes (scaling indices) $a_q$, with $q=2,3,4,5$.

\section{Two-dimensional case}
Let us investigate the presence  of non-statistical  fluctuations in
the (charged) multiplicity distributions depending on two variables
(like, e.g., the pseudorapidity and azimuthal angle). We present shortly
the way which allows  to formulate the relations  appropriate to this case.

First of all, we introduce the partitioning of the windows where both
quantities mentioned above were measured. The corresponding bins are
counted by the indices $m_1$ and $m_2$ where
$$  1~\leq ~ m_1~ \leq ~ M_1~~~{\rm and}~~~1 ~\leq ~ m_2~\leq M_2~.  $$
The number of charged particles, $n$, observed in the $(m_1, m_2, e)$-th
elementary cell is denoted by
\be
 n ~=~ n_{m_1, m_2, e} ~.
\ee
In this case the following notation is useful,
$$ {\bf I}_{m_1} ~\equiv~ {\bf I}_1~,~~~{\bf I}_{m_2}~ \equiv {\bf I}_2~;~~~
 {\bf B}_{m_1} ~\equiv {\bf B}_1 ~,~~~ {\bf B}_{m_2}~
\equiv~ {\bf B}_2~ $$
 and $ {\bf I}_e,~ {\bf B}_e$ represent the same averaging as in the 
preceding Section.

There are seven non-trivial types of the analysis; they can be
represented in the following form,  \\
\bea
&&{\bf [2 dim; 1]}~:~{\bf B}_e {\bf B}_2 {\bf I}_1,~~~~
{\bf [2 dim; 4]}~:~{\bf B}_e {\bf I}_2 {\bf I}_1~,      
  ~~   \nonumber  \\
&&{\bf [2 dim; 2]}~:~{\bf B}_e {\bf B}_1 {\bf I}_2,~~~~
{\bf [2 dim; 5]}~:~{\bf B}_2 {\bf I}_e {\bf I}_1~,    \nonumber  \\
&&{\bf [2 dim; 3]}~:~{\bf B}_2 {\bf B}_1 {\bf I}_e,~~~~  
{\bf [2 dim; 6]}~:~{\bf B}_1 {\bf I}_e {\bf I}_2~,   \nonumber    \\
&&~~~~~ ~~~~~ ~~~~~~{\bf [2 dim; 7]}~:~{\bf I}_e {\bf I}_2 {\bf I}_1 ~.  
\eea

\noindent
For instance, in the first case,
{\bf [2dim; 1]},
\bea
F_{m_1,m_2,e}(q) &=& \frac{{\cal F}(n;q)}{[{\bf I}_1 n]^q} ~,  \nonumber  \\
F_{m_2,e}(q) &=& {\bf I}_1 F_{m_1,m_2,e}(q) = \frac{{\bf I}_1 {\cal F}(n;q)}
{[{\bf I}_1 n]^q} ~ ,     \\
F(q) &=& {\bf B}_e {\bf B}_2 \frac{{\bf I}_1 {\cal F}(n;q)}{[{\bf I}_1 n]^q}
\propto f_q (M_1M_2)^{a_q}   \nonumber
\eea
where the multiplicity $n$ of the elementary cells is given by Eq. (12).

\section{Three-dimensional case}
If the question is to be answered whether the non-statistical fluctuations
are present in multiplicity data  depending simultaneously on three 
variables (like e.g. the pseudorapidity, azimuthal angle and transverse 
momentum), first of all the number of (charged) particles in every
elementary cell  should be specified. To this end, the windows in all three
variables are partitioned thereby involving the relations,
$1 \leq m_j \leq M_j$, with $j=1,2,3$. In this case, the multiplicity $n$
of the elementary cell is specified by
\be
n~~=~~ n_{{m_1}, {m_2}, {m_3},e}~~.
\ee
Introducing the notation analogous to that one  applied in the preceding
cases, namely
$$  {\bf I}_{m_j} \equiv {\bf I}_j~~~{\rm and}~~~ {\bf B}_{m_j}
\equiv {\bf B}_j~~~ j=1,2,3,    $$
we observe that in the present case the following fifteen possible
types of the analysis can be introduced,
\bea
{\bf [3dim; 1]}: {\bf B}_e {\bf B}_3 {\bf B}_2 {\bf I}_1~&,&~~
     {\bf [3dim; 3]}: {\bf B}_e {\bf B}_2 {\bf B}_1 {\bf I}_3   \nonumber  \\
{\bf [3dim; 2]}: {\bf B}_e {\bf B}_3 {\bf B}_1 {\bf I}_2~&,&~~
     {\bf [3dim; 4]}: {\bf B}_3 {\bf B}_2 {\bf B}_1 {\bf I}_e  \nonumber  \\
~~  \nonumber  \\   
{\bf [3dim; 5]}: {\bf B}_e {\bf B}_3 {\bf I}_2 {\bf I}_1~&,& ~~
     {\bf [3 dim; 8]}: {\bf B}_e {\bf B}_1 {\bf I}_3 {\bf I}_2   \nonumber  \\
{\bf [3dim; 6]}: {\bf B}_e {\bf B}_2 {\bf I}_3 {\bf I}_1~&,& ~~
     {\bf [3dim; 9]}:  {\bf B}_3 {\bf B}_1 {\bf I}_e {\bf I}_2     \\
{\bf [3dim; 7]}: {\bf B}_3 {\bf B}_2 {\bf I}_e {\bf I}_1~&,& ~~ 
     {\bf [3 dim; 10]}: {\bf B}_2 {\bf B}_1 {\bf I}_e {\bf I}_3   \nonumber  \\
~~  \nonumber    \\
{\bf [3dim; 11]}: {\bf B}_e {\bf I}_3 {\bf I}_2 {\bf I}_1~&,& ~~
     {\bf [3 dim; 13]}: {\bf B}_2 {\bf I}_e {\bf I}_3 {\bf I}_1   \nonumber  \\
{\bf [3dim; 12]}: {\bf B}_3 {\bf I}_e {\bf I}_2 {\bf I}_1~&,& ~~
     {\bf [3dim; 14]}: {\bf B}_1 {\bf I}_e {\bf I}_3 {\bf I}_2   \nonumber  \\
~~   \nonumber   \\
  {\bf [3dim; 15]}&:&{\bf I}_e {\bf I}_3 {\bf I}_2 {\bf I}_1 ~.   \nonumber  
\eea
In Eq. (16), the ${\bf B}_e$ and ${\bf I}_e$  have the same meaning
as in the preceding Sections. For instance,  the first type of the
analysis mentioned in Eq. (16) involves the factorial moments in the
$(m_1, m_2, m_3, e)$-th cell,
\be
F_{m_1,m_2,m_3,e}(q)= \frac{{\cal F}(n;q)}{[{\bf I}_1 n]^q}
\ee
as well as the full form of that moment together with
the corresponding scaling condition,
\be
    F(q) = {\bf B}_e {\bf B}_3 
{\bf B}_2 {\bf I}_1F_{m_1,m_2,m_3,e}(q) ~~\propto~~f_q (M_1 M_2 M_3)^{a_q}
\ee
where the multiplicity $n$ is given by Eq. (15). 

Let us add that  every type
of the analysis is characterized by its own set of the scaling
indices $a_q$ and the corresponding form of all kinds of the
associated moments (like the frequency, correlation, dispersion, etc.
moments); some more details in${^5}$.

\section{Dispersion moments}
It is well known that several
 multiplicity distributions observed in the past,
were satisfactorily described by continuous
statistical distributions inherent in different modelling ideas.
And quite often the associated moments, especially the corresponding
dispersion moments, represented a very apt tool which helped to
discriminate between more and less appropriate models. 
Now, it can be expected that especially at higher energies (available 
today as well as in the future)  the possible presence of the
intermittency should be taken into account.
    In this case, application of the continuous distributions is 
already not sufficiently adequate, compare in particular
e.g. Fig. 1 in${^6}$; that distribution, in${^6}$, representing one event,
is induced by non-self-similar processes and clearly reveals scaling 
property${^7}$ (also other authors deal with the individual events, for
instance${^8}$).
Application of the corresponding associated statistical moments (like 
e.g.  the appropriately introduced dispersion moments), might play again
a decisive role in the procedure leading to the discern between
the more or less reliable fractal model approaches. 
Moreover, the dispersion moments
involve also a piece of information about the presence of the empty 
bins. 

In particular, the first type of the analysis mentioned in the 
present paper, {\bf [1dim; 1]}, involves (let us call them 
``modified'') dispersion moments $\tilde{D}$ defined by the following 
way
\be
[\tilde{D}_q^q]_{me} = \left( \frac{n_{me}}{{\bf I}_{m'} n_{m'e}} - 1 \right)^q
\ee
and  $  \tilde{D}_q^q = {\bf B}_e {\bf I}_m [\tilde{D}_q^q]_{me}~.$
On the other hand, the ``standard'' dispersion moments $D$ are introduced by
\be
 [D_q^q]_{me} ~=~ (n_{me} - {\bf I}_{m'} n_{m' e})^q~,  
\ee
i.e. their relation 
with the modified $\tilde{D}$-dispersion moments acquires the
form
\be [\tilde{D}_q^q]_{me}~=~ \frac{1}{[{\bf I}_{m'} n_{m' e}]^q}
[D_q^q]_{me}  
\ee
and 
\be 
    \tilde{D}_q^q ~=~  
{\bf B}_e \frac{{\bf I}_m  [D_q^q]_{me}} {
    [{\bf I}_{m'} n_{m' e}]^q}
  ~=~ {\bf B}_e {\bf I}_m [\tilde{D}_q^q]_{me}~.
\ee
Moreover, in the case {\bf [2dim; 1]}:
\be
  [\tilde{D}_q^q]_{m_1, m_2, e} = \left( \frac{n_{m_1, m_2, e}}
{{\bf I}_{m'_1} n_{m'_1, m_2, e}}
-1 \right)^q~,  ~~ \tilde{D}_q^q = {\bf B}_e {\bf B}_{m_2} {\bf I}_{m_1} 
 [\tilde{D}_q^q]_{m_1, m_2, e}~;
\ee

\smallskip
\noindent
and {\bf [3dim; 1]}:
\bea
 [\tilde{D}_q^q]_{m_1,m_2,m_3,e} = \left( \frac{ n_{m_1, m_2, m_3, e}}
{{\bf I}_{m'_1} n_{m'_1, m_2, m_3, e}} - 1 
\right)^q~,     \nonumber   \\
  \tilde{D}_q^q =
{\bf B}_e {\bf B}_{m_3} {\bf B}_{m_2} {\bf I}_{m_1} \left(
\frac{n_{m_1, m_2, m_3, e}}
{{\bf I}_{m'_1} n_{m'_1, m_2, m_3, e}} - 1 \right)^q~.
\eea

We note that the multifractal analysis (of the type {\bf [1dim; 1]}) 
of data coming from the collisions of gold (at 10.6 A GeV/c)  with 
emulsion nuclei${^9}$ leads to the conclusion that the lower rank 
dispersion moments (essentially of the form  of Eq. (19))
reveal scaling property with sufficient accuracy${^{10}}$.

 Analogous procedure can be applied also in higher dimensional cases.

\section{Conclusions}
   In the present contribution, investigating presence of the anomalous 
fluctuations, a natural and systematic extension of the one-dimensional 
approach  to arbitrary, say, d-dimensional case is proposed; it involves   
$$\sum_{j=0}^d \left( \begin{array}{cc}
d+1  \\ j \end{array}  \right) ~=~ 
 2^{d+1} - 1 $$ 
non-trivial
types of the analysis. The aim is to apply such a systematic approach
to a same set of the data and to look for the regularities
embedded in the scaling indices which belong to the same branch of the
analyses. Hopefully, the regularities which would be observed there 
might lead at least to the construction of the corresponding model
equations.

   Especially, it is expected that a systematic application of the 
non-trivial types of the analysis to several samples of events will allow 
   (i) to recognize    which type of the analysis is the most
suitable one for detecting the presence of the concrete type of the
anomalous fluctuations,
   (ii) to introduce an appropriate cathegorization of the anomalous 
phenomena, 
    (iii) to ascribe a concrete numerical value to the parameters
which enter theoretical expressions applied to the
description of the experimental data,
and, eventually,
   (iv) to formulate a sufficiently reliable (and still missing) 
theory (or at
least model) of the anomalous phenomena.

\end{document}